\documentclass[11pt]{article}
\pdfoutput=1

\usepackage{geometry}
\makeatletter

\usepackage{amsfonts,amsmath,amssymb,bm,eucal}
\usepackage[english]{babel}
\usepackage[latin1]{inputenc}
\usepackage[T1]{fontenc}
\usepackage{ae,aecompl}

\usepackage{graphicx,graphics,color,psfrag,subfigure,epsfig,epstopdf}
\usepackage{hyperref}
\usepackage[comma,numbers,sort&compress]{natbib}

\title{
\vspace*{-2cm}
\begin{flushright}
\normalsize{BARI-TH/2011-648}
\end{flushright}
\vspace*{1.5cm}
\textbf{Temperature and quark density effects \\ on the chiral condensate: an AdS/QCD study}\\ ~\\
\author{P.~Colangelo$^a$, F.~Giannuzzi$^{a,b}$, S.~Nicotri$^{a,b}$, V.~Tangorra$^b$\\
~\\ 
\normalsize\emph{$^a$Istituto Nazionale di Fisica Nucleare, Sezione di Bari, Italy}\\
\normalsize\emph{$^b$Dipartimento di Fisica, Universit\`a degli Studi di Bari, Italy}
}}

\date{}

\begin{document}

\pdfbookmark[1]{Temperature and quark density effects on the chiral condensate: an AdS/QCD study}{Temperature and quark density effects on the chiral condensate: an AdS/QCD study}
\maketitle

\begin{abstract}
We investigate the dependence of the chiral condensate $\langle\bar qq\rangle$ on the temperature and quark density using the soft-wall holographic model of QCD, adopting geometries with black holes at finite temperature and 
quark chemical potential $\mu$.
We find that, for  $\mu$ below a critical value, increasing the temperature the condensate decreases and vanishes at a temperature $\tilde T\simeq 210$ MeV (at $\mu=0$).
An analogous behaviour is observed increasing the chemical potential at fixed temperature.
These results agree with the findings obtained by other methods.
We also comment on the robustness of the results if geometries not involving black holes are adopted at low temperature, and an Hawking-Page transition is implemented.
\end{abstract}

\vspace{1cm}

The strong coupling regime of quantum Yang-Mills theories, where perturbative methods are not helpful, represents a problem requiring continuous investigations.
This is particularly true for QCD, so rich of important soft effects the full understanding of which is still lacking. 
The knowledge of the dynamics of the QCD vacuum would shed light on aspects of primary importance, such as color confinement and spontaneous chiral symmetry breaking.  \cite{Shifman:2010jp}

For $N_f$ massless quarks quantum chromodynamics shows a global $U(N_f)_L\otimes U(N_f)_R=SU(N_f)_L\otimes SU(N_f)_R\otimes U(1)_B\otimes U(1)_A$ symmetry, whose $SU(N_f)_L\otimes SU(N_f)_R$ part, at zero temperature and vanishing quark chemical potential, is broken down to vector $SU(N_f)_V$ (the baryon $U(1)_B$ symmetry remains exact, the axial $U(1)_A$ is anomalous).
The spontaneous breaking of this chiral symmetry is at the origin of effects such as the presence of Goldstone bosons (for $N_f$=2 they are the pions) and the large mass splitting between, e.g., light vector and axial-vector mesons, and is signaled by a non-vanishing vacuum expectation value of the $\bar q q = \bar q_R q_L+\bar q_L q_R$ bilinear quark operator.

If the temperature $T$ or the quark density (or the quark chemical potential $\mu$) are increased, QCD is supposed to undergo a transition to a phase in which the chiral $SU(N_f)_L\otimes SU(N_f)_R$ symmetry is restored. This produces changes of the hadron properties, and has the consequence 
that the quark-antiquark condensate $\langle \bar qq\rangle$ vanishes \cite{BraunMunzinger:2009zz}.

The behaviour of $\langle\bar qq\rangle$ versus temperature and quark chemical potential is therefore of fundamental interest, and indeed lattice QCD and effective field theory analyses have been employed to determine such a dependence. 
In particular, results have been obtained by lattice QCD as far as the $T$ dependence is concerned \cite{Bazavov:2009zn}, while in the case of the quark density dependence, the problem connected to the non-positive definite fermionic determinant in the numerical evaluation of the discretized QCD partition function has not been fully solved \cite{Fodor:2001au}.
Effective theories of QCD have also been used to establish such dependences, with a particular role played by chiral perturbation theory \cite{Gerber:1988tt}, and by the Nambu-Jona Lasinio (NJL) model together with its Polyakov-NJL extension \cite{Buballa:2003qv}.

A method to access the nonperturbative regime of QCD has been developed in the recent past years, and is inspired by the Anti de Sitter/Conformal Field Theory (AdS/CFT) or gauge/gravity correspondence conjecture \cite{maldacena1,gubser-klebanov-polyakov,Witten:1998qj}.
It is denoted as holographic QCD (or AdS/QCD), and assumes the existence of a duality between large-$N_c$ QCD and a higher dimensional weakly coupled semiclassical field theory formulated on a curved AdS-like spacetime. 
Using perturbative methods in the weakly coupled theory, one can obtain information about the dual strongly coupled QCD, since the correspondence implies that the partition functions of the two theories coincide, under suitable boundary conditions \cite{Witten:1998qj}, and that the Minkowski spacetime on which the gauge theory is defined can be identified with the boundary of the AdS space.

Our purpose is to study the dependence of the chiral condensate on the temperature and quark density in a holographic framework denoted as soft-wall model \cite{sw,andreev1} (investigations in different holographic approaches can be found, e.g., in \cite{Kobayashi:2006sb,Iatrakis:2010jb}).
The soft-wall model is defined in a non-dynamical AdS$_5$ spacetime;  finite temperature and quark density effects are described considering geometries with black holes.
In the case of finite temperature, the line element in the gravity model is chosen as the AdS-Schwarzschild (AdS/BH),
\begin{equation}\label{metric}
 ds^2=g_{AB}dx^A dx^B=\frac{R^2}{z^2} \left(f(z)\, dt^2-d\bar x^2-\frac{dz^2}{f(z)}\right) 
\end{equation}
with
\begin{equation}\label{metric1}
f(z)=1-\left(\frac{z}{z_h}\right)^4 \,\,\, .
\end{equation}
$R$ is the AdS$_5$ radius and $z$ the fifth (holographic) coordinate ($\epsilon \le z$, with $\epsilon \to 0^+$).
$z_h$ is the position along $z$ of the horizon of a black hole, and it is related to the Hawking temperature $T$ by the condition 
\begin{equation}\label{HT}
 T=\frac{1}{4\pi}\,\left|\frac{df}{dz}\right|_{z=z_h} \,\,\, ;
\end{equation}
$T$ is identified with the temperature of the QCD boundary theory.
Notice that we use capital latin indices for $5D$ coordinates, and Greek indices for $4D$ coordinates.
With this choice of the metric we are assuming that the AdS/BH geometry is the stablest one, i.e. the lowest-action gravity configuration for all values of the temperature; we shall come to this point in the following.

The soft-wall model involves a complex scalar ($p=0$-form) field $X(x,z)$ \cite{sw}, whose mass is fixed by the AdS/CFT prescription: $m_5^2 R^2=(\Delta-p)(\Delta+p-4)$ \cite{Witten:1998qj}. 
This field is dual to the $\Delta=3$ dimension QCD $\bar qq$ operator. 
The gravity theory also includes $5D$ gauge fields, $L_M$ and $R_M$, dual to the $4D$ $\Delta=3$ left-and right-handed currents $j_{L}^\mu=\bar q_{L}\gamma^\mu q_{L}$ and $j_{R}^\mu=\bar q_{R}\gamma^\mu q_{R}$ associated to the global $SU(N_f)_{L}$ and $SU(N_f)_{R}$ symmetries, respectively: hence, the global chiral symmetry of the QCD boundary theory is gauged in the bulk. 
The $5D$ meson-action defining the holographic model reads (we set to 1 the AdS$_5$ radius $R$) \cite{sw,Colangelo:2008us}
\begin{equation}\label{action}
S=\frac{1}{k_{YM}} \int d^4x\int_0^{z_h}dz\,\,e^{-\phi(z)}\sqrt{g}\,\mbox{Tr}\left[|DX|^2- m_5^2 |X|^2 -\frac{1}{4g_5^2}\left(F_L^2+F_R^2\right)\right]\,,
\end{equation}
where $g$ is the determinant of the metric \eqref{metric} and $F_{L,R}$ are the field strenghts $F_{L}^{MN}=F_{L}^{MNa}T^a=\partial^M { L}^N-\partial^N {L}^M-i \left[ {L}^M, { L}^N \right]$, and an analogous expression for $F_R$. 
The constants $k_{YM}$ and $g_5^2$ have been fixed in \cite{Erlich:2005qh,Colangelo:2008us}: $\displaystyle k_{YM}=\frac{16\pi^2}{N_c}$ and $\displaystyle g_5^2=\frac{3}{4}$.
The dilaton $\phi(z)$ is a background field introduced to account for the breaking of conformal invariance in the IR, with the functional form $\phi(z)=c^2z^2$ chosen to produce confinement \cite{zakharovpot} and linear Regge trajectories for light hadrons \cite{sw}. 
The scale parameter $c$ is fixed from the mass of the $\rho$ meson to the value $c=m_\rho/2=389$~MeV \cite{sw} (in the following we set $c=1$ and express the dimensionful quantities in units of $c$).

The field $X$ is in the bifundamental representation of $SU(N_f)_L\otimes SU(N_f)_R$, while the gauge fields are in the adjoint representation of the corresponding Lie algebras: $L_M=L_M^a T_L^a$ and $R_M=R_M^a T_R^a$, with $T_{L,R}^a$ the generators of $SU(N_f)_{L,R}$\footnote{$L_M$ and $R_M$ are in the singlet representation of $SU(N_f)_R$ and $SU(N_f)_L$, respectively.}; the trace is taken over gauge indices.
The covariant derivative acting on $X$ is defined as $D_M X=\partial_M X-iL_M X+iX R_M$, and it is the way in which the two fields $L_M$ and $R_M$ are coupled to each other. 
Therefore, the scalar sector of the theory, i.e. $X$, is responsible of chiral symmetry breaking in the model (for $X=0$ chiral symmetry is restored) \cite{DaRold:2005zs}, \cite{Son:2003et}.
The fluctuations around the vacuum expectation value of $X$ describe light scalar mesons \cite{Colangelo:2008us,sw}.
To complete the model, a Chern-Simons term should be added to describe the chiral anomaly \cite{Colangelo:2011xk}; the dual of another $\Delta=3$ spin one operator ${\CMcal O}^{\mu\nu}=\bar q\sigma^{\mu\nu}q$ has also been proposed to be included in the gravity action \cite{Domokos:2011dn}.
From now on, we consider the case of two quark flavours, putting $N_f=2$.
The result can be easily generalised to other values of $N_f$.

In the action \eqref{action} the integration over the holographic coordinate $z$ is extended up to the black-hole horizon $z_h$, which represents an IR bound and sets the size of the bulk. 
We assume that the behaviour of $\langle\bar qq\rangle$ is due to the interaction with the geometry and the dilaton.
This means that we only consider linearized equations for the field $X$, ignoring its coupling to the gauge fields, as well as other contributions coming from a possible potential term $V(X)$ which could be added to the action Eq.\eqref{action} \cite{sw}.

Defining the scalar field as $X(x,z)=X_0(z)\mathbf{1}_{N_f} e^{i\pi(x,z)}$ (hence neglecting the fluctuations around the $X_0$ configuration, while $\pi(x,z)=\pi^a(x,z)T^a$ represents chiral fields), where $\mathbf{1}_{N_f}$ is the $N_f\times N_f$ identity matrix, we can identify $X_0(z)$ as the v.e.v. describing the dynamics of the condensate $\langle\bar qq\rangle$: as it can be inferred from dimensional analysis, the coefficient of the term proportional to $z^3$ in the expansion of $X_0$ for $z\to 0$ is indeed proportional to $\langle\bar qq\rangle$ \cite{Klebanov:1999tb},\cite{Erlich:2005qh}.
To determine such a coefficient, we work out the equation of motion for $X_0(z)$ from the action \eqref{action}:
\begin{equation}\label{eomscalar}
 X_0''(z,z_h)- \frac{2 z^2 f(z)+4-f(z)}{z f(z)} X_0'(z,z_h)+\frac{3}{z^2 f(z)} X_0(z,z_h)=0 \,
\end{equation}
in which we have explicitely indicated the dependence of $X_0$ on the position $z_h$ of the horizon, hence on the temperature. The primes denote derivatives with respect to $z$.
At $T=0$ (i.e. for $z_h\to\infty$ and $f(z)=1$) this equation can be solved analytically and has only one regular solution in the (large $z$) IR region,
\begin{equation}\label{tricomi}
 X_{0}^{T=0}(z)=\frac{m_q\sqrt{\pi}}{2}\, z\, U\left(\frac{1}{2},0,z^2\right)\,
\end{equation}
with $U$ the Tricomi confluent hypergeometric function. The solution should also be multiplied by a factor describing the scaling of $X_0$ with $N_c$ \cite{Cherman:2008eh,Colangelo:2011xk}, which however is not relevant for the present analysis.
The coefficient $m_q$ is identified with the quark mass by the condition that $X_0/z$ must tend to the source of the $\bar qq$ operator in QCD as $z\to0$ \cite{Witten:1998qj}. 
Eq.~\eqref{eomscalar} also admits another linearly independent solution, which is singular for $z\to\infty$; it corresponds to a divergent on-shell action (\ref{action}) and therefore it is discarded.

In Eq.(\ref{tricomi}) the chiral condensate, the coefficient of the $z^3$ term in the expansion of $X(z)$ at small $z$, and the quark mass, the coefficient of the linear $z$ term, are proportional, a relation which does not hold in QCD. 
Indeed, expanding \eqref{tricomi} for $z\to0$ we get the asymptotic form
\begin{equation}
 X_{0}^{T=0}(z) \xrightarrow[]{z\to0} m_q\, z+m_q\, \left(\frac{1+\gamma_E}{2}-\log(2)\right)z^3+m_q\,z^3 \log(z)+ {\CMcal O}(z^5)\,. 
\end{equation}
This result is a drawback of the soft-wall model, and it does not allow to separately describe explicit and spontaneous chiral symmetry breaking. 
In other holographic approaches, namely the hard-wall model \cite{Erlich:2005qh}, the coefficients of the $z$ and $z^3$ terms in the small $z$ expansion of $X_0$ are not related to each other; however, in those cases the chiral condensate is an external input and cannot be extracted from the dynamics of the gravity system (studies of temperature effects in the hard-wall holographic model can be found in \cite{Ghoroku:2005kg,Kim:2007em}).
A way out in the soft-wall framework includes the possibility of adding a potential term to \eqref{action} \cite{sw,Gherghetta:2009ac}, together with other solutions \cite{Afonin:2011ff}.
In spite of this, it is possible to have hints on the dependence of $\langle\bar qq\rangle$ on the temperature at $T\neq0$. 

Near the black-hole horizon ($z\sim z_h$) Eq.(\ref{eomscalar}) admits two solutions, a logarithmically divergent solution and a regular solution which behaves as
\begin{equation}\label{solzh}
 X_0(z,z_h)\xrightarrow[]{z\to z_h} 1-\frac{3}{4}\,\left(1-\frac{z}{z_h}\right)-\frac{3 (5-8z_h^2)}{64}\,\left(1-\frac{z}{z_h}\right)^2+{\CMcal O}\left((z_h-z)^3\right)\,.
\end{equation}
This regular solution for $T \to 0$ continuously reproduces the zero temperature expression of $X_0$ given by Eq.(\ref{tricomi}), therefore it can be used to fix the profile of $X_0$ close to $z_h$.
On the other hand, at low values of $z$, the two independent solutions have asymptotic form
\begin{eqnarray}
 X_{01}(z,z_h) & \xrightarrow[]{z\to 0} & z+\frac{3}{4}z^3+z^3\log(z)+ {\CMcal O}(z^5) \label{lowzX}\\
 X_{02}(z,z_h) & \xrightarrow[]{z\to 0} & z^3+ {\CMcal O}(z^5)\,, \label{lowzX1}
\end{eqnarray}
with the term proportional to $z^3\log z$ having the same coefficient of the linear $z$ term. 
The dependence on $z_h$ (hence on the temperature $T$) does not enter the low-$z$ behaviour up to ${\CMcal O}(z^5)$. 
The general solution has then the form $X_0=A\,X_{01}+B\,X_{02}$ and the condition $\displaystyle \frac{X_0}{z}\Big|_{z \to 0}=m_q$ fixes the coefficient of (\ref{lowzX}) to $A=m_q$.
Therefore, the low-$z$ dependence of $X_0$ is
\begin{eqnarray}\label{lowzsigma}
 X(z,z_h) & \xrightarrow[]{z\to 0} & A\,X_{01}(z,z_h)+B\,X_{02}(z,z_h)\\
& \xrightarrow[]{z\to 0} & m_q z+m_q z^3 \log(z)+\left(\frac{3}{4}\,m_q+B\right) z^3 \nonumber\\
& = & m_q z+m_q z^3 \log(z)+\sigma z^3\,,\nonumber
\end{eqnarray}
where $\sigma=\frac{3}{4}\,m_q+B$ is proportional to the chiral condensate $\langle\bar qq\rangle$ through a temperature-independent coefficient; it is possible to find this relation from a matching condition with QCD of a one-point correlation function of a scalar operator \cite{Colangelo:2011xk} (see appendix).
Such a proportionality relation ensures a correct $N_c$ scaling of the field $X_0$ ($\sim N_c^0$), the quark mass ($\sim N_c^0$) and the chiral condensate ($\sim N_c$) \cite{Cherman:2008eh}.
In the soft-wall model, the coefficient $\sigma(T)$ must be fixed through a suitable IR boundary condition.
As such boundary condition, we require the general solution to match the behaviour \eqref{solzh} as $z\to z_h$, which ensures regularity and continuity with the $T=0$ case.
To do this, we use a numerical shooting method, where the shooting parameter is the coefficient $\sigma$ in \eqref{lowzsigma}. 
The value of this parameter turns out to be different for different values of temperature, and applying the method for different $T$'s we obtain the dependence of $\sigma$ (and, so, on the chiral condensate $\langle\bar qq\rangle$) on $T$ in this model.
$m_q$ is the bare quark mass and it is an external parameter in the soft-wall model \cite{sw}; we keep it fixed with respect to temperature, as e.g. in \cite{Kobayashi:2006sb,Iatrakis:2010jb} and from this point on, we set $m_q=1$ for simplicity.

\begin{figure}[b!]
 \centering
 \includegraphics[scale=0.7]{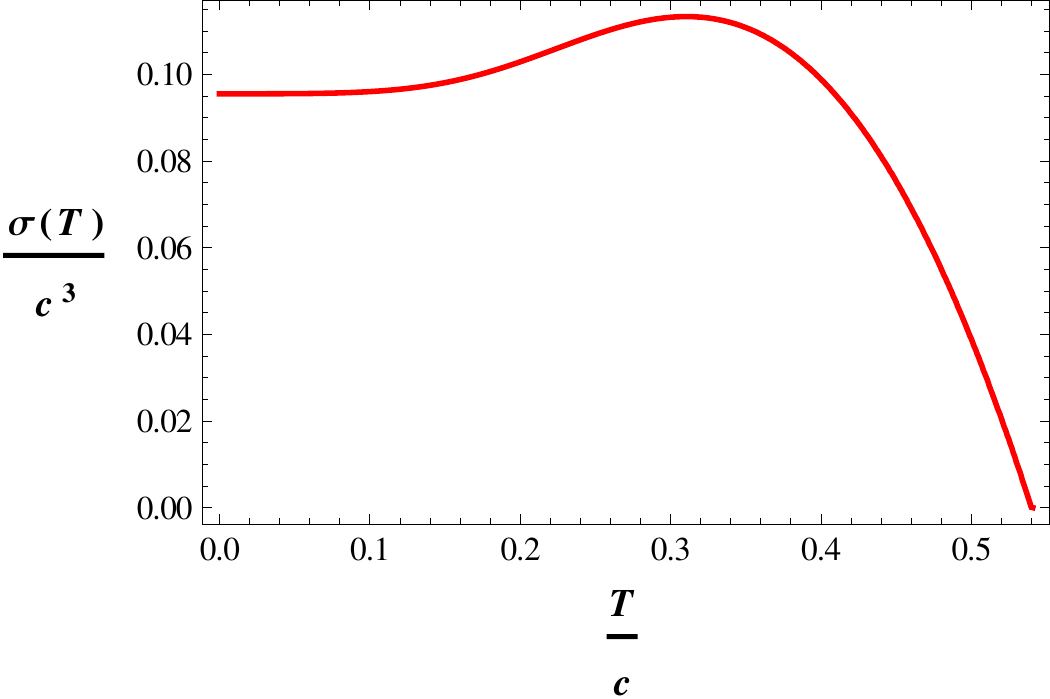}
\caption{Dependence of $\sigma$ in Eq. \eqref{lowzsigma} on the temperature $T$.
Both quantities are expressed in units of the scale $c$.}
 \label{fig:sigma}
\end{figure}

The resulting $\sigma(T)$ is drawn in Fig.~\ref{fig:sigma}.
Starting from the value at $T=0$ and increasing the temperature, after a slight increase to a local maximum at $T/c\sim0.31$, the condensate drops to $\sigma=0$ at $\tilde T/c\sim 0.54$. 
Numerically, this temperature corresponds to $\tilde T\sim210$~MeV.
This behaviour is similar to what is commonly expected: for example, in some finite temperature QCD sum rule analyses, the condensate profile $\displaystyle \sigma(T)=\sigma_0\left(1-\left(\frac{T}{T_c}\right)^\alpha\right)$ is used (at least close to the critical temperature $T_c$, with $\sigma_0$ the value at $T=0$) \cite{Dominguez:2007ic}.
Analogous results are obtained by Schwinger-Dyson calculations \cite{Blank:2010bz} and in models of QCD \cite{Halasz:1998qr,Huang:2011mx}. 
Since for $T\geqslant T_c$ the chiral condensate vanishes, for massless quarks one has chiral symmetry restoration and $T_c$ is the temperature at which the chiral phase transition occurs, with $\alpha$ a critical exponent. 
Lattice QCD analyses also find a decreasing chiral condensate, which vanishes at $T \simeq 180-200$ MeV \cite{Bazavov:2009zn}.
A summary of the results obtained through different lattice approaches can be found in \cite{Borsanyi:2010bp}.

In the soft-wall model chiral symmetry is never restored, since the quark mass remains different from zero; explicit symmetry breaking occurs and the chiral limit cannot be properly investigated.
On the other side, the fact that the behaviour of $\sigma$ for $T\leqslant\tilde T$ is similar to the one expected in QCD for $T\leqslant T_c$, is an interesting feature in view also of possible modifications aimed at an improved description of chiral symmetry within this model.
Extending $T$ to higher values, $\sigma$ becomes negative, therefore $\tilde T$ represents the maximum temperature below which the model can be suitably applied to hadrons. This temperature is higher than the one where 
melting is observed in the spectral functions of the lightest hadrons \cite{spectra-function}.

Let us now  comment on the  geometry assumed to be dual to finite temperature QCD. 
In frameworks in which a dynamical geometry is considered (in particular in the case of the gravity dual of ${\CMcal N}=4$~SYM), it has been observed that two different solutions of the Einstein equations in the bulk have to be taken to describe the finite temperature field theory on the boundary,  the AdS/BH and the thermal-AdS (th-AdS). The latter is characterized by AdS metric and compact periodic Euclidean time dimension, with size proportional to $1/T$. Depending on the temperature, the two geometries correspond to different stationary points of the gravitational action, th-AdS being the stablest solution at low temperature, AdS/BH the stablest one at high temperature. A Hawking-Page transition between  th-AdS and AdS/BH is interpreted as the gravity dual of the QCD deconfinement transition. 

In the soft-wall model, assuming that the th-AdS+dilaton and AdS/BH+dilaton geometries correspond to exact solutions of  gravity equations and considering the on-shell gravity action,  the existence of a Hawking-Page transition has been envisaged \cite{Herzog:2006ra}, with a deconfinement temperature $T_{HP}\simeq\frac{1}{\pi}\frac{c}{0.647}\simeq 0.49\,c$ \footnote{A critical temperature $T_C\simeq 210$ MeV, associated to the deconfinement transition, has been obtained in a  soft-wall model with AdS/BH geometry and no Hawking-Page transition, analyzing the static quark-antiquark pair  potential  \cite{Andreev:2006eh}.}. 
If the th-AdS and the AdS/BH geometries were adopted in the present calculation of the chiral condensate, following the method discussed above, the chiral condensate would remain constant up to $T=T_{HP}$, and then would follow the curve in Fig.\ref{fig:sigma} for higher temperatures. 
The small increase of the condensate with temperature shown in Fig.\ref{fig:sigma}  (not observed on the lattice and in chiral perturbation theory) would be absent, and the profile for $\sigma$ at $T=T_{HP}$ would be discontinuous.
However, the assumption that the geometry and the dilaton field effectively result from the solution of a coupled gravity-dilaton equation is crucial for the existence of the Hawking-Page transition \cite{Herzog:2006ra}.
On the other hand, the soft-wall model  adopted here is  phenomenologically constructed, the dilaton profile being suitably chosen to reproduce empirically observed features of the dual theory, and  its  dynamical origin is not known.
Even standing these limitations, the soft-wall model can be considered a minimal benchmark in the  attempt to get information on important features of the QCD vacuum.

To investigate the dependence of the chiral condensate on the quark density, we need to introduce the quark chemical potential in the holographic framework.
Finite quark density effects can be described by adding to the action, in the QCD generating functional, the term $\int d^4x\,\mu q^\dagger q$, where $q^\dagger q=\bar q\gamma^0q$ is the quark number operator and $\mu$ the quark chemical potential. 
Therefore, $\mu$ can be considered as the boundary value of the time component of a $U(1)$ gauge field in the bulk theory.
The geometry of the bulk coming from the interaction between a space-time with negative cosmological constant (the AdS space) and a $U(1)$ gauge field results in an AdS/Reissner-Nordstr\"om (AdS/RN) space, an AdS space containing a charged spherically symmetric black hole.
For vanishing spatial components of the $U(1)$ gauge field $A_i(z)=0$ ($i=1,2,3,z$), the small $z$ expansion of the time component $A_0(z)$ can be written as
\begin{equation}\label{muparameters0}
 A_0(z)=\mu-\kappa q z^2 
\end{equation}
where $q$ is the charge of the black-hole and $\kappa$ a dimensionless coefficient which can be considered a parameter of the model \cite{Colangelo:2010pe}.
$\kappa$ scales as $\sqrt{N_c}$, and its (model dependent \cite{Jo:2009xr}) numerical value represents a parameter useful to phenomenologically implement subleading effects in the large $N_c$ limit. 
The results in the following refer to $k=1$.
The line element of the AdS/RN gravity theory has the same form as in \eqref{metric}, with the black-hole factor modified to
\begin{equation}
 f(z)=1-(1+Q^2)\,\left(\frac{z}{z_h}\right)^4+Q^2\,\left(\frac{z}{z_h}\right)^6 \qquad\quad 0<Q^2<2\,,
\end{equation}
where $Q$ is $Q=qz_h^3$ \cite{Colangelo:2010pe}.
The quantities $Q$ and $z_h$ are connected to the temperature $T$, defined by (\ref{HT}), and to the chemical potential $\mu$:
\begin{eqnarray}\label{muparameters}
T & = & \frac{1}{\pi z_h}\biggl(1-\frac{Q^2}{2}\biggr)\nonumber\\
\mu & = & \kappa\,\frac{Q}{z_h} \,\,\, 
\end{eqnarray}
with the second equation coming from the condition that $A_0$ vanishes at the horizon.

\begin{figure}[ht]
 \centering
 \vspace{0.5cm}
 \subfigure[]{\includegraphics[scale=0.6]{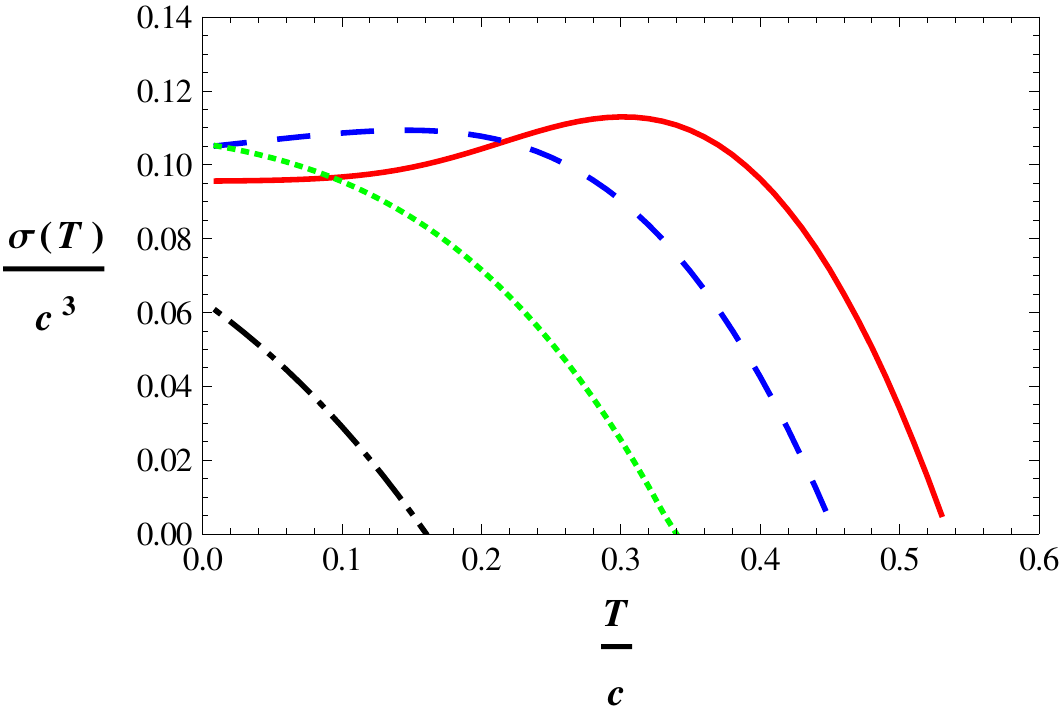}\label{fig:sigmaTparamMu}}
\hspace{0.5cm}
 \subfigure[]{\includegraphics[scale=0.6]{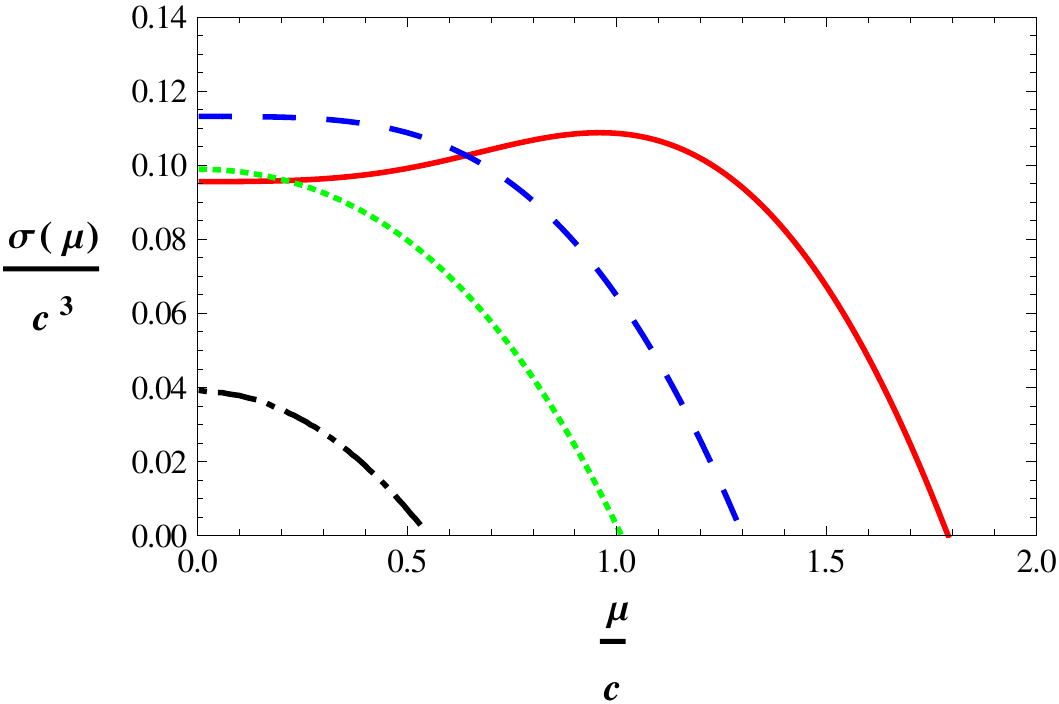}\label{fig:sigmaMuparamT}}
\caption{Chiral condensate versus the temperature $T$ at different values of the quark chemical potential $\mu$ (a), and versus $\mu$ at different values of $T$ (b).
In (a) the curves correspond to $\mu/c=0.02$ (plain, red), $0.8$ (blue, dashed), $1.2$ (green, dotted) and $1.6$ (black, dot-dashed).
In (b) the curves correspond to $T/c=0.05$ (plain, red), $0.3$ (blue, dashed), $0.4$ (green, dotted) and $0.5$ (black, dot-dashed). The parameter $\kappa$ is set to $\kappa=1$.}
\label{fig:sigmaparam}
\end{figure}

Following the same procedure outlined above to solve \eqref{eomscalar}, we find that as $z\sim0$, the two linearly independent solutions follow Eqs.(\ref{lowzX},\ref{lowzX1}). 
On the other hand, near the horizon, the regular solution now behaves as
\begin{eqnarray}\label{solzhFD}
 X(z,z_h) & \xrightarrow[]{z\to z_h} & 1-\frac{3}{4}\,\left(1-\frac{z}{z_h}\right)\nonumber\\
&&+\frac{3 \left(5-2Q^2+4(Q^2-2)z_h^2\right)}{16(Q^2-4)}\,\left(1-\frac{z}{z_h}\right)^2+{\CMcal O}((z_h-z)^3)\,.
\end{eqnarray}
We numerically solve the equation of motion \eqref{eomscalar} tuning $\sigma$ in Eq.~\eqref{lowzsigma} to obtain the asymptotic dependence \eqref{solzhFD} at $z\sim z_h$, and the results of the calculation are collected in Fig.~\ref{fig:sigmaparam}.
The behaviour of $\sigma$ versus $T$ for different values of chemical potential $\mu$ is plotted in \ref{fig:sigmaTparamMu}: for all the considered values of the chemical potential, increasing the temperature the chiral condensate drops to zero at values of the temperature that depend (decrease) on $\mu$. 
An analogous behaviour is found exchanging $T$ with $\mu$, as shown in Fig.~\ref{fig:sigmaMuparamT}. 
These dependences agree with the results obtained by different methods, for example in approaches based on the composite operator formalism \cite{Barducci:1993bh}. 
The physical value of $\mu$ for which, at low temperatures, the chiral condensate vanishes depends on the parameter $\kappa$ in (\ref{muparameters0}): using the value $\kappa=1/2$ fixed (in a different model) in \cite{Colangelo:2010pe}, it corresponds to $\tilde\mu\simeq350$ MeV.

Analogously to the $\mu=0$ case, also at finite density, if one considers a dynamical geometry and dilaton, it is necessary to establish which is the stablest geometry for each value of  $T$ and $\mu$.
This problem has been studied in top-down approaches \cite{Kobayashi:2006sb} and in the bottom-up hard-wall model   \cite{Jo:2009xr,Sin:2007ze}.
In \cite{Kobayashi:2006sb} it is shown that for $\mu>0$ the thermal-AdS solution is inconsistent, and the geometry always contains a black-hole.
In particular, a first order phase transition between two black-hole geometries occurs at a critical $T_c(\mu)$, up to a certain value of $\mu$.
In  the hard-wall model  it is found that a transition between a ``charged-th-AdS'' solution and the AdS/RN one could take place \cite{Jo:2009xr,Sin:2007ze}.
In our phenomenological model the geometry and the dilaton are fixed background  configurations,  the existence of transitions between different geometries being related to the unknown solutions of gravity equations:  for this reason we  only consider AdS/RN for all values of $T$ and $\mu$.

\begin{figure}[b!]
 \centering
 \includegraphics[scale=0.65]{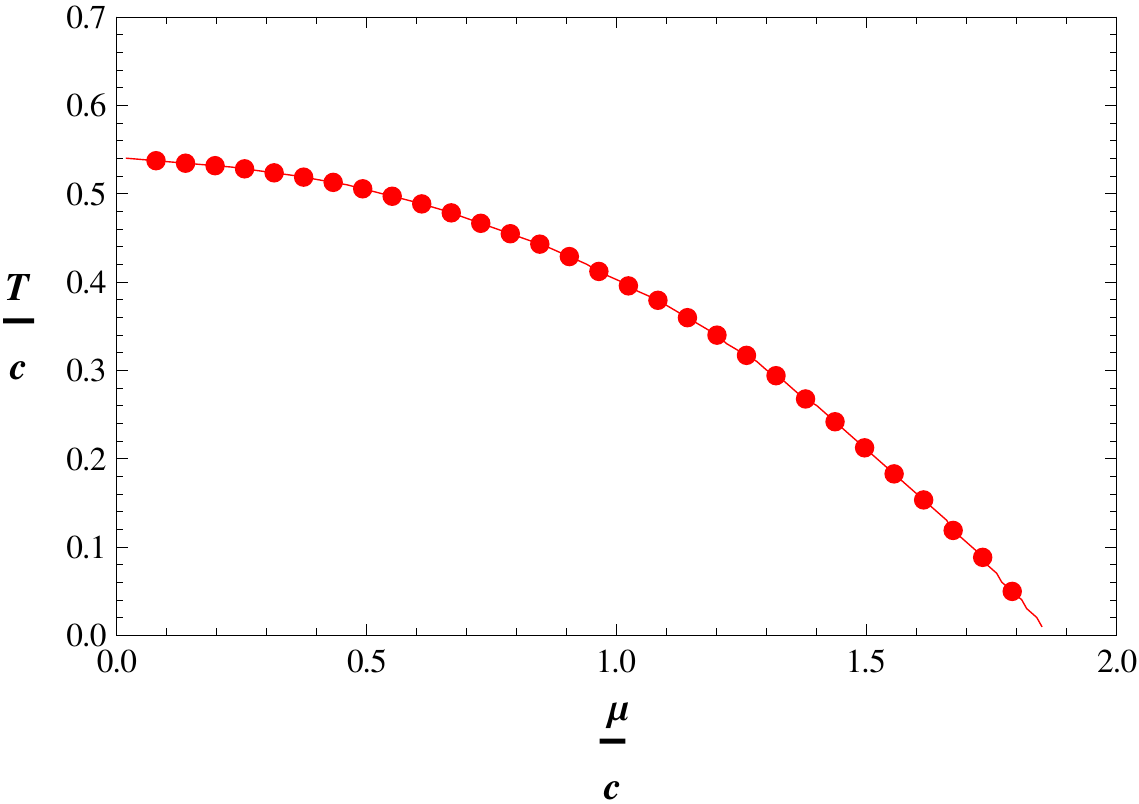}
\caption{Line in the $T-\mu$ plane where $\sigma$ vanishes. The parameter $\kappa$ is set to $\kappa=1$.}
 \label{fig:phd}
\end{figure}

In Fig.~\ref{fig:phd} we plot the values of the temperature $\tilde T$ where $\sigma$ vanishes, versus the chemical potential $\mu$. 
It is similar to the one expected in the case of transitions to a phase with restored chiral symmetry: in the case of the soft-wall, the curve depicted in Fig.~\ref{fig:phd} indicates the region of $T$ and $\mu$ in which the model can be used to describe hadrons.

To conclude, studying $\langle\bar qq\rangle \propto\sigma$ at finite temperature and quark density using the soft-wall holographic model of QCD and geometries with black holes, we have found a behaviour of the condensate versus the temperature in qualitative agreement with the results of other approaches.
In particular, increasing the temperature $\sigma(T)$, after a slight increase up to a maximum, drops to zero at a value $\tilde T\simeq 210$~MeV which is close to other determinations. 
For finite quark chemical potential the behaviour is analogous, and the line in the temperature/chemical potential plane in which $\sigma$ vanishes is similar to the line envisaged for QCD.
Such results  encourage the hope that a dual theory of QCD can be  formulated.

\vspace*{1.0cm}
\noindent {\bf Acknowledgments.} \\
 We thank F.~De Fazio and F.~Jugeau for helpful discussions and collaboration in the early stage of this work.
 This work is supported in part by the Italian MIUR PRIN 2009.

\appendix

\pdfbookmark[1]{Appendix}{Appendix}
\section*{Appendix}

A proportionality between the coefficient $\sigma(T)$ and the chiral condensate is found here, showing that all the temperature dependence of $\langle\bar q q\rangle$ is contained in $\sigma(T)$, while the coefficient of proportionality is temperature-independent and ensures a correct scaling of quantities with the number of colours.
The chiral condensate can be computed within holographic approaches through the relation
\begin{equation}
\langle\bar qq\rangle=\lim_{m_q\to 0}\sqrt{\frac{2}{n_f}}\,\left\langle J_S^{0}\right\rangle \,, 
\end{equation}
being $\left\langle J_S^{0}\right\rangle$ the vacuum expectation value of the quark scalar current.
In order to evaluate this quantity, let us consider the on-shell action for the scalar field
\begin{equation}
 \CMcal{S}=\frac{1}{k_{YM}}\int d^4x\, \left. \frac{e^{-\phi(z)}\,f(z)}{z^3}\, \mbox{Tr}\left[ X(x,z)\, \partial_z X(x,z)\right] \right|_{z=0} \,,
\end{equation}
where
\begin{equation}
X(x,z)=\left(X_0(z)\, {\bf 1}+S(x,z)\right)e^{2i\pi}\,.
\end{equation}
$X_0(z)$ is the expectation value we have considered in this paper, $\pi$ is the chiral field, and $S(x,z)$ is the fluctuation describing scalar mesons \cite{Colangelo:2008us}:
\begin{equation}
S(x,z)=\int d^4y \, S(x-y,z)\, S_0^a(y)\, T^a\,,
\end{equation}
$S(x-y,z)$ being the bulk-to-boundary propagator and $S_0^a$ the source, according to the holographic dictionary \cite{Witten:1998qj}; $T^a\, (a=1,..8)$ are the Gell Mann matrices and $T^0={\bf 1}/\sqrt{2n_f}$.
Then in the holographic model the relation holds:
\begin{equation}
 \left\langle J_S^{0}(x)\right\rangle = -\left. \frac{\delta\CMcal{S}}{\delta S_0^0}\right|_{S_0^0=0}\,,
\end{equation}
and, in the Fourier space, defining $\tilde S(q,z)=\int d^4x\, e^{-iq\cdot x}S(x,z)$, the vev of the scalar current reads:
\begin{equation}\label{holovev}
\left\langle J_S^{0}\right\rangle=-\frac{\sqrt{2n_f}}{2\, k_{YM}}\left. \frac{e^{-\phi(z)}\,f(z)}{z^3}\, \left[ X_0(z)\, \tilde S'(0,z)+ X_0'(z)\, \tilde S(0,z)\right]\right|_{z=0}\,,
\end{equation}
where the prime indicates a derivative with respect to $z$.
At finite temperature and for low values of the fifth coordinate $z$, the scalar field $\tilde S(q,z)$ can be written as:
\begin{equation}\label{scalS}
\tilde S(q,z) \xrightarrow[]{z\to 0} z + A(T)\, z^3 \log z+B(T) z^3 + \CMcal
{O}(z^5) 
\end{equation}
where we have used the boundary condition $\tilde S(q,z)/z\xrightarrow[]{z\to 0} 1$, and $A(T)$ and $B(T)$ are coefficients we do not need to specify for the scope of the calculation.
The low-$z$ behaviour of $X_0$ is
\begin{equation}\label{scalX}
X_0(z) \xrightarrow[]{z\to 0} m_q\, z + m_q\, z^3 \log z+\sigma(T) z^3 + \CMcal
{O}(z^5) \,,
\end{equation} 
where only $\sigma(T)$ depends on the temperature, as shown in the paper.
Plugging \eqref{scalS} and \eqref{scalX} into \eqref{holovev}, we find the proportionality relation between the coefficient $\sigma(T)$ and the chiral condensate\footnote{The factor here is different from the one found in \cite{Colangelo:2011xk} due to a different definition of the scalar field $X_0$.}
\begin{equation}\label{propcoeff}
\langle\bar qq\rangle = -\frac{4}{k_{YM}} \sigma(T)=-\frac{N_c}{4\pi^2}\sigma(T)\,,
\end{equation}
provided $\sigma(T)$ is finite in this limit.
This result shows that the temperature dependence of the chiral condensate is only governed by $\sigma(T)$, the proportionality coefficient $N_c/(4\pi^2)$ not depending on $T$.
It is worth stressing that such a coefficient ensures the correct scaling of all the quantities involved in Eq.~\eqref{scalX} with $N_c$. 
This solves the problem of the scaling of the vev with $N_c$ pointed out in \cite{Cherman:2008eh}, since all the coefficients in the expansion \eqref{scalX} scale as $N_c^0$, while $\langle\bar qq\rangle$ scales as $N_c$.


\begin{thebibliography}{99}
\bibitem{Afonin:2011ff}
It has also been proposed to restrict the validity of the model to the IR region, in which case $m_q$ should be interpreted as a constituent quark mass:
  S.~S.~Afonin,
  Phys.\ Rev.\ C\ {\bf 83} (2011) 048202.
\end{thebibliography}
\end{document}